# An empirical model of fleet modernization: on the relationship between market concentration and innovation adoption in the Brazilian airline industry

Alessandro V. M. Oliveira✢

Thiago Caliari

Rodolfo R. Narcizo

This version: 12 December 2020

**Abstract**

The modernization of an airline's fleet can reduce its operating costs, improve the perceived quality of service offered to passengers, and mitigate emissions. The present paper investigates the market incentives that airlines have to adopt technological innovation from manufacturers by acquiring new generation aircraft. We develop an econometric model of fleet modernization in the Brazilian commercial aviation over two decades. We examine the hypothesis of an inverted-U relationship between market concentration and fleet modernization and find evidence that both the extremes of competition and concentration may inhibit innovation adoption by carriers. We find limited evidence associating either hubbing activity or low-cost carriers with the more intense introduction of new types of aircraft models and variants in the industry. Finally, our results suggest that energy cost rises may provoke boosts in fleet modernization in the long term, with carriers possibly targeting more eco-efficient operations up to two years after an upsurge in fuel price.

✢ Corresponding author. Email address: alessandro@ita.br.

▪ Affiliations: Center for Airline Economics, Aeronautics Institute of Technology, Brazil.

▪ Acknowledgments: The first author wishes to thank the São Paulo Research Foundation (FAPESP)–grants n. 2013/14914-4 and 2015/19444-1; National Council for Scientific and Technological Development (CNPq)–grants n. 301654/2013-1, n. 301344/2017-5; The third author thanks the Coordination for the Improvement of Higher Education Personnel (CAPES)–Finance Code 001. The authors wish to thank Maurício Morales, Paulo Ivo Queiroz, Cláudio Jorge P. Alves, José Guerreiro Fregnani, Cristian Vieira dos Reis, Luiz André Gordo, Vitor Caixeta Santos, Paula Guimarães, Rogéria Arantes. All remaining errors are ours.

# 1. Introduction

To survive in a market with fierce cost competition and subject to increasingly intense environmental pressure, an airline needs to make adequate strategic planning of its fleet. One alternative is an effective fleet rollover policy that replaces older aircraft with newer units of the aircraft types already available either in the manufacturers' menu list or in the secondary aircraft markets. A more aggressive strategy may consider actively modernizing the fleet with the new generations of aircraft launched by manufacturers. Although a modern fleet's operation can lead to higher costs related to aircraft lease payments and amortization, it may also bring higher fuel cost savings, as the new models are typically more energy efficient.[1] Also, operating modern airplanes may enhance carriers' frequent flier passengers' loyalty and willingness to pay since more advanced aircraft may be associated with greater in-flight comfort and perceived quality. As put by Clark (2017), airlines nowadays stress the aircraft's ability to contribute to its brand as a critical product attribute. So, ultimately, the acquisition of new generation airplanes may constitute a win-win situation that brings competitive advantages of demand and costs, besides possibly allowing faster mitigation of the airline's emissions. In contrast, however, there is still much debate about whether the pace of the air transport sector's innovation trajectory has been sufficient to meet reasonable emission-mitigation goals (Peeters et al., 2016).

The present paper aims to empirically investigate the incentives carriers have to adopt manufacturers' technological innovations by acquiring new-generation aircraft. We propose an econometric model of fleet modernization determinants in the Brazilian commercial aviation over two decades. In Brazil, air fleets are among the youngest in Latin America, with major airlines Latam, Gol, and Azul among Boeing and Airbus's main customers with orders for the most modern models since their launch.[2] Additionally, competition has notably intensified in this industry after deregulation, with market concentration declining 15% since the mid-2000s.[3]

We examine the role of airline market competition as a driver for the strategic fleet planning and assignment decision-making process. More specifically, we empirically test the hypothesis of an "inverted U-shaped" relationship between market concentration and innovation adoption in the airline industry. Aghion et al. (2005) suggest an inverted-U relationship to model firms' innovation behavior under different market settings. According to the authors, in industries subject to a few technologically asymmetric firms' higher monopoly power, market concentration would be positively associated with innovation, and therefore the *Schumpeterian effect* would prevail. In contrast, in industries marked by neck-and-neck competition, with firms having similar technological levels, market concentration would be negatively associated with innovation, as the soothing of competition would disincentive the "*escape-competition* effect" of innovative behavior. With the possibility of an inverted U-shape, both the extremes of competition and concentration may inhibit firms' innovation adoption. As far as we are concerned, we are the first to inspect such non-linearities in the innovation adoption behavior of transportation markets. We also contribute to the literature by proposing an index of innovation adoption based on a fleet modernization indicator constructed from aircraft tail number-specific information. We consider the date of the introduction of each aircraft model type and variant by manufacturers as the reference for our fleet modernization indicator.

---

[1] For example, concerning the Boeing 737 MAX and Airbus A320neo families, the manufacturers claim the associated models are up to 20% more fuel-efficient than the respective previous generations. Source: "Environment Report - Build a Better Planet" (2014), available at www.boeing.com; and "A320neo Family sets new standards with 20% reduced fuel burn" (2014), available at www.airbus.com.

[2] See, for example, "*TAM Airlines becomes first A350 XWB operator from the Americas,*" December 18, 2015, available at www.airbus.com; "*Gol Airlines becomes first South American operator to order Leap-1B-powered 737 MAX,*" October 1, 2012, available at www.cfmaeroengines.com; and "*Azul receives the first Airbus A330neo in the Americas,*" May 13, 2019, available at www.airbus.com.

[3] Source: National Civil Aviation Agency's (ANAC) Air Transport Statistical Database, with own calculations.

In our proposed econometric setting, we account for the unobserved effects of fleet assignment by considering thousands of possible nuisance parameters that control for demand, cost, strategic and technology-related factors. These confounders may cause omitted-variables bias in the estimation of the fleet modernization equations. We utilize the post-double LASSO model (PDS-LASSO, Belloni, et al., 2012) to select the empirical models' relevant focus and control variables. We cross-check the results with estimations of Bayesian model-averaging (BMA, Leamer, 1978) and weighted-average least squares (WALS, Magnus, Powell, and Prüfer, 2010) models.

Our results provide support for the inverted-U shaped hypothesis in the innovation adoption behavior of airlines. Additionally, we find limited evidence that hubbing activity or low-cost carriers may be associated with the more intense introduction of new types of aircraft models and variants in the industry. Finally, our results suggest that energy costs increases may boost fleet modernization in the long term, with carriers possibly targeting more eco-efficient operations up to two years after an upsurge in fuel price.

This paper is organized as follows. Section 2 provides a discussion of the literature on fleet management and innovation adoption in the airline industry. Section 3 presents the application and discussion of the empirical approach. Section 4 presents the estimation results, and Section 5 presents the conclusions.

## 2. Strategic fleet management and the incentives for innovation adoption in the airline industry

### *2.1. Airline fleet modernization and innovation adoption*

One of the main challenges of an airline's strategic planning concerns the management of its fleet. According to Holloway (2008), this planning stage consists of adapting the productive capacity to the expected future demand–a task that requires decisions regarding the acquisition of new aircraft. Airlines generally purchase planes to replace the existing capacity or, in the case of an increase of traffic, network, or market share, to expand their supply of seats in the market. Fleet management also involves projecting the operating costs of the renewed fleet and the possibilities for generating revenue. Therefore, strategic fleet planning aims to guide the balanced introduction of the airline's capacity into the market and enable competitive advantages over rivals.

Given that the delivery of a new aircraft from the original equipment manufacturers–OEMs–can take years, proper fleet planning must be accomplished within a medium-to-long-time horizon, which involves considering several uncertainties. It must consider the information set available decision-making time and incorporate the firm's expectations about factors such as overall economic growth, the evolution of competition, and trends in the adopted business model. However, the risks associated with each decision under consideration can be high, as in the airlines that purchased the Boeing 737 MAX model and subsequently had their plans interrupted with the aircraft's global grounding after the 2018 2019 accidents.[4] Airlines resort to sequential reviews of fleet planning across time to alleviate the risks associated with the decision-making process, shaping it according to the new market conditions' materialization. First, they can manage their order books with the OEMs to balance between firmly ordered and optioned aircraft–Clark (2007).[5] They can also perform tactical adjustments to the aircraft delivery schedule using deferral rights stipulated when signing the purchase contract.[6] Other possibilities include utilizing leased aircraft with a rolling spread of return options (Holloway, 2008).

The management of an airline's fleet involves preventing aircraft aging–fleet rollover–and making decisions regarding the insertion of technological innovation in the market by acquiring more modern aircraft. Traditionally, air transport literature is relatively little concerned with issues related to technical progress in air transport.[7] The shortage of studies is probably related to the fact that more disruptive technological innovations–such as the jet passenger aircraft in the 1960s–have not been persistent in the sector (Brueckner & Pai, 2009). Franke (2007), on the other hand, argues that in aviation, innovation can still be implemented in three areas: (i) through new business models; (ii) through advanced passenger segmentation; and (iii) through new technologies or more efficient operations–in this case, by the introduction of new generation aircraft types. The author argues that these three types of innovation can collaborate decisively in strengthening the potential for generating new revenues and cutting operating costs necessary to survive in such a competitive market.

---

[4] See "*Boeing: global grounding of 737 Max will cost company more than $1bn*", April 24, 2019, available at theguardian.com.

[5] Clark (2007) describes, however, that whereas firmly-ordered aircraft is contractually binding, optioned aircraft are subject to separate negotiations between the airline and the OEM by the time of the future conversion of the options to actual orders.

[6] Examples can be found at "*LATAM Airlines Group to defer 22 aircraft deliveries through 2015*", August 21, 2013, available at centreforaviation.com, and "*Azul postpones A330neo delivery until high season 2019*", November 13, 2018, available at centreforaviation.com.

Albers, et al. (2020) analyze the effects of recent disruptive innovation in air transport–introducing the business model of long-haul and low-cost airlines, such as the Asian AirAsia X, Scoot, or Jetstar. The authors describe that this new business model results from aircraft technology's progress, the growing experience in generating ancillary revenue and cargo operations, and the liberalization of international markets. Brueckner & Pai (2009) investigate the impact of the emergence of regional jets on airline service standards and quality of service in the US market since the late 1990s. The authors' results suggest that, contrary to what was expected ex-ante, the technological innovation brought about by the operations with regional jets do not generate point-to-point routes in new, less dense markets, but in markets with demographic characteristics similar to those then existing. They point to evidence that carriers assign regional jets to provide services on many new hub-and-spoke routes, having replaced the discontinued jet and turboprop service, and supplemented the continuing jet service.

Regarding the technical progress allowed by incorporating new airplanes into a fleet, Holloway (2008) discusses two aircraft innovativeness elements. The first is the generation of the aircraft, given that the younger generations naturally have improvements in design, reliability, ease of maintenance and may present disruptive innovations, such as with more efficient engines–rotation times improved, modular maintenance facility–, winglets, use of lightweight composite materials in the fuselage, instead of aluminum, among others. The second factor is the age of a specific aircraft in an airline's fleet, usually related to the number of flight hours and cycles recorded, with a natural generation of corrosion and fatigue and a consequent increase in engine' and components maintenance costs.

We will refer to the inclusion of new generation aircraft by the airlines, and its actual assignment to a given route, as "fleet modernization" in that market. This definition is consistent with the term that is commonly found in aircraft purchase news. For example, in 2017, in the United States, American Airlines announced an investment of US $ 20 billion in aircraft over four years, which its CEO qualified as being "*the most aggressive fleet modernization in the history of the industry*[8] In Canada, Air Canada announced in early 2020 the continuation of its fleet modernization program, with its first Airbus A220.[9] In Brazil, Gol and Azul airlines announced fleet modernization plans in 2018 and 2019, respectively–the former increased its 737 MAX order, and the latter announced the arrival of additional Airbus A320neo and Embraer E195-E2.[10] It is important to note that airlines can perform fleet modernization by simply discarding older generation aircraft models from their fleet.

---

[7] An exception is the studies involving cost function estimation, as in Caves, Christensen & Tretheway (1984), which take technical progress into account as an unmeasurable variable, through time trend or time effect dummies in the case of data in time series.

For the strategic considerations regarding fleet modernization, the airline should analyze OEMs' aircraft's broad portfolio. The air transport literature has recently focused on the diversity of existing aircraft models, using fleet standardization indices, as in Zou, Yu & Dresner (2015) and Narcizo, Oliveira & Dresner (2020). Zou, Yu & Dresner (2015), for example, find evidence that airlines can improve profitability by standardizing their fleet at the level of aircraft families, but standardization at the aircraft model level can lower profits.

The literature has recently utilized OEMs' distinction between aircraft "families" and "models." Kilpi (2007) describes that "family" is the set of aircraft in which the differences between the flight equipment are limited, related to the length of the fuselage, payload capacity, and engines–they may have common flight and cabin crews, as well as common spare components. Another possible level of detail for fleet analysis is the "aircraft model," a jargon commonly used in the industry to refer to the OEM's product–such as, for example, the Boeing models "737-700" and "737-800", and Airbus "A320ceo" and "A320neo." In our analysis, we consider the difference between aircraft model types and their variants. The model variants are also known as "model versions" in the industry. Table 1 shows an illustration using some of the Airbus A320 series products. For example, if the airline chooses to purchase an Airbus model of type A321neo, it can choose, for example, the -251N, -253N, -271N, and 272N variants, launched between 2016 and 2017, or the -252N variant, launched in

---

[8] "*American Airlines Bets Big on Aggressive Fleet Modernization*," March 20, 2017, available on aviationoutlook.com.

[9] "*Air Canada Celebrates the Arrival of its First Airbus A220, Continuing its Fleet Modernization Program*", January 15, 2020, available on www.staralliance.com.

[10] "*GOL Airlines Increases Boeing 737 MAX Order to 135 Aircraft; Converts 30 Orders to Larger MAX*", July 16, 2018, available on ir.voegol.com.br; "*Azul moves forward with fleet modernization*," November 8, 2019, available on www.aerotime.aero. Other airline news that utilizes the term "fleet modernization: "*American Airlines Continues Modernization Launching New aacargo.com*," June 9, 2020, available at www.caasint.com; "*Southwest Airlines looks to modernize fleet*," Apr 4, 2018, available at www.smartbrief.com; "*Airbus A380 marks start to fleet modernization for British Airways*," July 3, 2013, available at br.reuters.com; "*Lufthansa Cargo Accelerates Fleet Modernization*," Nov 7, 2019, available at www.aviationpros.com; "*Wizz Air to slow fleet modernization over next three years*," June 3, 2020, available at www.flightglobal.com; "*FedEx Fleet Modernization Continues with Introduction of New Boeing 767F*," October 24, 2019, available at www.caasint.com.

2019.[11] It is not proper to state that a variant of a model launched more recently is necessarily more modern than the variants launched earlier, because in some cases, the different coding has some purely commercial purpose of the manufacturer, or it may have many relevant operational characteristics in common, without changes significant among them. However, we cannot rule out the possibility that these most recent variants contain improvements introduced over the years, resulting from customers' continuous feedback, thus presenting some innovative content compared to previously released versions. In some cases, essential attributes such as maximum range, maximum take-off weight, type of engine, and winglets' existence constitute a true differential between aircraft model variants.

**Table 1–Aircraft model types and variants: an illustration**

| Model Type | Model Variant | Start | Engine | Thrust (kN) | Model Type | Model Variant | Start | Engine | Thrust (kN) |
|---|---|---|---|---|---|---|---|---|---|
| A320ceo | A320-111 | Feb-87 | CFM56-5A1 | 111 | A321ceo | A321-131 | Mar-93 | IAE V2530-A5 | 140 |
| | A320-211 | Jun-88 | CFM56-5A1 | 111 | | A321-111 | May-93 | CFM56-5B1 | 133 |
| | A320-231 | Oct-88 | IAEV2500-A1 | 111 | | A321-112 | Mar-94 | CFM56-5B2 | 138 |
| | A320-212 | Sep-90 | CFM56-5A3 | 118 | | A321-231 | Dec-96 | IAE V2533-A5 | 147 |
| | A320-232 | Apr-92 | IAE V2527-A5 | 120 | | A321-211 | Mar-97 | CFM56-5B3/P | 147 |
| | A320-214 | Mar-95 | CFM56-5B4/P | 120 | | A321-232 | Sep-01 | IAE V2530-A5 | 140 |
| | A320-233 | Oct-95 | IAE V2527E-A5 | 120 | | | | | |
| | | | | | A321neo | A321-251N | Feb-16 | CFM LEAP-1A32 | 143 |
| | | | | | | A321-271N | Mar-16 | PW1133G-JM | 147 |
| A320neo | A320-271N | Sep-14 | PW1127G-JM | 120 | | A321-253N | Apr-17 | CFM LEAP-1A33 | 143 |
| | A320-251N | May-15 | CFM LEAP-1A26 | 121 | | A321-272N | Aug-17 | PW1130G-JM | 147 |
| | A320-252N | Nov-19 | CFM LEAP-1A24 | 107 | | A321-252N | Oct-19 | CFM LEAP-1A30 | 143 |

Sources: websites planelogger.com, and bgspotters.net ("Deciphering the Airbus codes"). "Start" denotes the first flight date.

We raise the issue that carriers' modernization by carriers should not be characterized as a real innovative behavior in airline markets. More precisely, we claim that the acquisition of new generation aircraft is more appropriately defined as "technological innovation adoption." The adoption of new technologies is linked to the characteristics of innovation (Rogers, 1995), the organizational aspects of the adopter (Damanpour, 1991), the features of the supplier (Frambach & Schillewaert, 2002), and the social and economic characteristics (Frambach & Schillewaert, 2002; Rogers, 1995). Given (i) that the most significant source of investments in R&D in aviation are the OEMs, such as Boeing and Airbus; and (ii) that there is no substitutability of products (only from suppliers), there is a strong influence of OEMs in the technological adoption of the market, defining trajectories and dependencies. Airlines, being consumers of OEM technology products, are responsible for deciding when and at what intensity to introduce new aircraft models and their variants on the market. The conceptual difference between "innovation" and "innovation adoption" is essential, given that the adoption of technological innovation, in general, does not share the same characteristics as the innovation itself, as the guarantee of the exclusive appropriability of the technology introduced. On the contrary, in commercial aviation, in general, the announcement of the acquisition of a more modern aircraft is made publicly, and it is not possible to prevent rivals from engaging in a race to acquire the same technology, or similar, within the same future horizon and thus mitigating the competitive advantage gains from the innovative airline.

---

[11] There are also the -251NX, -252NX, -253NX, -271NX e -27N variants.

## *2.2. Competition, concentration, and the market incentives to innovate*

The literature discussion about the relationship between competition and innovation is intense and typically confined to inter-industry studies. From the Schumpeterian point of view, the increase in innovative activity is closely related to monopoly power–Baldwin and Scott (1987) and Kamien and Schwartz (1982)–, in a market setting in which low market uncertainty, along with economies of scale and scope, are necessary conditions for the increase in the returns on innovative investment–Chandler Jr. (1994). Aghion et al. (2005) denote the evidence from Industrial Organization studies that predict a negative relationship between competition and innovation as the *Schumpeterian effect*. On the other hand, they point out that many empirical studies in the area find the opposite: a positive relationship between competition and innovation, denoting the *escape-competition effect*. The authors reconcile the two strands of studies by finding evidence of an "inverted-U shaped" relationship-between competition and innovation. With an inverted-U relationship, it would be possible to explain companies' behavior both in asymmetric markets with leader-follower competition patterns–more subject to the Schumpeterian effect–and symmetrical markets with "neck-and-neck" competition–more subject to the escape-competition effect.

The relationship between competition and innovation may also be associated with proximity to incumbent companies' technological frontier. Aghion et al. (2009) describe that firms close to the technological frontier tend to react to new firms' entry with the intensification of innovative activities, as they have a greater survival success with the realization of innovation. The authors call it the *escape-entry effect* phenomenon. According to the authors, this argument is a causal relationship from Schumpeter's theory of growth, similar to the escape-competition effect. In sum, market symmetry and proximity to the technological frontier tend to intensify the adoption of innovation in the face of competitive pressures, and these two characteristics would provide micro-foundations for the inverted-U relationship.

With these studies, the literature on the relationship between competition and innovation has intensified, giving rise to a series of empirical studies (Peneder, 2012). Bérubé et al. (2012), through a database of manufacturing companies in Canada (2000-2005 period), find evidence that the competitive intensity is positively related to the company's spending on research and development (R&D). However, they find that this relationship decreases as the distance from the sectoral technological frontier increases; in short, competition positively impacts innovation when carried out between equals–corroborating the relationship between market symmetry and proximity to the technological frontier. Analyzing the Dutch industry, Polder & Veldhuizen (2012) reached similar results, supporting the inverted-U relationship between competition and innovation. However, they point out that the dissemination of technology (narrowing the technological gap) within the industry acts as an inhibiting effect, making the competition-innovation relationship negative. Hashmi (2013) analyzes the inverted-U relationship for the case of publicly traded manufacturing companies in the United States of America. The results point to a slightly negative relationship, robust to a series of alternative specifications, but with the theoretical model's modification, indicate suggest both negative and "inverted U" relationships. As an explanation, the authors assume that the UK manufacturing industry (the basis for Aghion et al. (2005) has a more neck-and-neck competitive pattern than its US counterparts. Askenazy et al. (2013) and Beneito et al. (2017) also find evidence of an inverted-U relationship using inter-industry studies.

### 2.3. Airline fleet modernization and inverted-U relationship

Signals from the competitive environment can be critical to an airline's market incentives to modernize its fleet. With more modern aircraft and proper aircraft assignment across its multiple markets,[12] combined with effective marketing efforts, the airline can enhance its consumer loyalty and fortify its market positions from its rivals' price movements. In Brazil, all major carriers have been engaging in brand-strengthening campaigns in which fleet modernization is a key attribute. For example, in 2016, Latam announced its new A320neo aircraft's operation with the nose art painted on its fuselage as "The first A320neo of South America." In late 2019, Azul Airlines highlighted in its marketing campaign that it was the first Brazilian company to massively operate Embraer's E-Jets, painting the country's flag on the side of the fuselage of its brand-new Embraer E195-E2 aircraft. These examples illustrate the association between an airline's fleet modernization decisions and its marketing efforts.[13]

We consider Aghion et al. (2005) framework as a possible way to model innovation adoption in airline markets. To our knowledge, this is the first attempt to investigate the inverted-U hypothesis with an intra-industry study and the first econometric study of the determinants of innovation adoption in transportation markets. In the specific case of aviation, the inverted-U relationship may represent the market circumstances that make airlines decide whether or not to adopt a technological innovation through fleet modernization. In general, the degree of technological asymmetry between airlines tends to be low, given that most incumbents may gain access to the latest products from OEMs. However, asymmetries in access to capital markets and proper financing of aircraft acquisitions can lead to technological disparities between large and medium-sized carriers over time.

---

[12] See Jansen & Perez (2016).

[13] Source: "*Latam reveals the first image of the A320neo*", July, 12, 2016. available in Portuguese at www.panrotas.com.br. "*GOL' changes' name of Boeing 737 MAX 8 to 737-8*", April 1, 2019, avaialble in Portuguese at www.aeroin.net; and "*Azul will have an Embraer E195-E2 plane with the Brazilian flag on the fuselage*", October 27, 2019, available in Portuguese at www.aeroin.net. An example that goes in the opposite direction, namely, when a technological attribute of the aircraft is detrimental to the airline's image, is related to Gol Airlines' 2019 marketing campaign. In that campaign, Gol renamed it Boeing "737 MAX 8" to "737-8", after the two air accidents that forced the grounding of that aircraft model.

We employ a line of reasoning adapted from Aghion et al. (2005) to motivate the inverted-U relationship in the air transport sector. We perform such adaptation given that we have an intra-industry study–instead of inter-industry analysis–, with markets in which direct investment in R&D is not an everyday possibility but the adoption of technological innovation from suppliers. Thus, in commercial aviation, we could observe an inverted U-shape curve when comparing different city-pairs, or the same city-pairs in different periods, regarding the companies' fleet modernization behavior. Consider two cases: low and high market concentration, as measured by the HHI. Firstly, on routes or periods of low HHI, an intensification of competition may trigger intense price competition and bring financial fragility, or even bankruptcy, for some carriers. With lower prices and a potential loss of market share, the incremental benefit of purchasing a new aircraft may decrease. In this case, disincentives for fleet modernization could arise since the financial condition for new aircraft acquisitions is impaired or that capital costs are not justified at that time. The companies' strategic fleet planning problem would, therefore, be subject to a stricter financial constraint. Secondly, in city-pairs or periods where the measured HHI is high, increased competition could provoke the opposite behavior of airlines, that is, to encourage the adoption of technological innovations with the introduction of new-generation aircraft. This phenomenon would be equivalent to the escape-competition effect, where competition increases the incremental profits from innovation. The new acquisitions would be now more viable, possibly due to carriers' greater financial strength given the soothed competition in the market and greater market power.

In our case, we consider a panel of air travel markets, in which the instances of city-pairs and periods with high and low market concentration are possible outcomes. The most commonly observed data, however, is associated with moderate levels of market concentration. In such in-between markets, it may not be possible to easily predict the effect of an intensification of competition on carriers' fleet modernization behavior. It may be that more competition provokes higher innovation adoption in some cases but lower in other cases. In such data behavior, we may observe patterns that are consistent with the inverted-U relationship hypothesis. Our primary research objective is to design an empirical model of fleet modernization to test that hypothesis formally.

# 3. Research design

## 3.1. Application

We develop an application considering the domestic airline industry in Brazil from 2004 to 2018. In Table 2, it is possible to see some of the critical market and operations figures of the market, focusing on the short- and long-term dynamics since the 2008 global oil price shock. By the end of this period, the four largest Brazilian airlines were Avianca, Azul, Gol, and Latam. In 2018, these airlines operated 481 airplanes, and the mean fleet age was 7.6 years.

**Table 2–Short- and long-run evolution of airline fleet age in Brazil after the 2008 oil price shock**

| Years | Market Dynamics | | Mean Fleet Age (Years) | | | Past Information Set | |
|---|---|---|---|---|---|---|---|
| | Airline Capacity (ASK) | Fleet Size (Count) | Aircraft Level | Variant Level | Model Level | Jet Fuel Price (2 yrs lag) | Market Concentration (2 yrs lag) |
| **2008** | 75.40 | 377 | 8.5 | 13.9 | 19.5 | 3.03 | 0.54 |
| **2010** | 102.78 | 432 | 7.8 | 13.4 | 19.0 | 3.48 | 0.57 |
| **2018** | 117.94 | 481 | 7.6 | 14.4 | 22.1 | 1.76 | 0.46 |
| **% Var** | | | | | | | |
| 2008/10 | *36.3%* | *14.6%* | *-8.6%* | *-3.5%* | *-2.3%* | *14.8%* | *5.4%* |
| 2008/18 | *56.4%* | *27.6%* | *-10.7%* | *3.6%* | *13.6%* | *-42.0%* | *-15.0%* |

*Sources: National Civil Aviation Agency's (ANAC) Active Scheduled Flight Historical Data Series—VRA; Air Transport Statistical Database; Brazilian Aeronautical Registry—RAB; National Agency for Petroleum, Natural Gas and Biofuels—ANP; websites planelogger.com, airfleets.net, jetphotos.com, and aviacaopaulista.com; state-specific legislation and online media news; "Market Concentration" denotes the Herfindahl-Hirschman Index (HHI); except for "Fleet Size," all measures extracted at the city-pair level; figures computed with authors' calculations.*

We can see in Table 2 that the industry has grown remarkably, from 75.40 billion ASKs in 2008 to 117.94 billion in 2018–an increase of 56.4%. As a result of the airlines' fleet rollover policies, there has been an apparent fleet renewal movement, from a mean fleet age measured at the aircraft level of 8.5 (2008) to 7.6 years (2018). Most of such decline occurred in the two years after the oil price shock of 2008. In Table 1, we present the jet fuel price lagged by two years to inspect such effect. Considering the decision-making horizon for aircraft acquisition as a period of two years, the mean fleet age observed in 2008 may be an outcome of the fleet planning circumstances of 2006. In 2006, the price of jet fuel was R$ 3.03–in inflation-adjusted local currency–, and soared to R$ 3.48 reais in 2008, a 14.8% increase. Therefore, we suggest that such a decrease may be related to the drop in mean aircraft age in the period. In Table 1, it is also possible to see that market concentration has increased by 5.4% from 2006 to 2008. Table 3 allows a more in-depth analysis of the existing flight equipment and its associated degree of fleet modernization in the industry. It presents the evolution of the top 10 most operated model variants in the sample period's Brazilian domestic air transport market.

Table 3–Top 10 most operated model variants

| | 2004 | | | 2008 | | | 2012 | | | 2016 | | |
|---|---|---|---|---|---|---|---|---|---|---|---|---|
| | Model Variant | Flights Share | Years since Launch | Model Variant | Flights Share | Years since Launch | Model Variant | Flights Share | Years since Launch | Model Variant | Flights Share | Years since Launch |
| 1 | A320-232 | 11.9% | 12.2 | A320-232 | 18.1% | 16.2 | B737-8EH(WL) | 15.8% | 6.4 | B737-8EH(WL) | 21.1% | 10.4 |
| 2 | F100 | 7.9% | 17.6 | B737-8EH(WL) | 11.1% | 2.5 | A320-214 | 10.9% | 17.3 | E190-200IGW | 11.2% | 9.6 |
| 3 | B737-76N | 7.6% | 5.8 | A320-214 | 9.8% | 13.3 | A320-232 | 8.3% | 20.2 | ATR 72-600 | 9.4% | 5.1 |
| 4 | A319-132 | 7.5% | 5.8 | A319-132 | 7.6% | 9.8 | A319-132 | 7.2% | 13.8 | A320-214 | 8.3% | 21.3 |
| 5 | EMB-120RT | 5.1% | 18.9 | B737-322 | 5.1% | 21.7 | E190-200IGW | 4.9% | 5.6 | E190-200LR | 8.0% | 10.2 |
| 6 | B737-33A | 5.1% | 17.8 | B737-76N | 4.7% | 9.8 | ATR 72-600 | 3.6% | 1.2 | A320-214(WL) | 7.1% | 14.6 |
| 7 | B737-2A1 | 4.3% | 35.1 | ATR 42-300 | 3.7% | 23.2 | E190-200LR | 3.6% | 6.2 | A319-132 | 5.4% | 17.8 |
| 8 | B737-53A | 3.4% | 14.1 | F100 | 3.4% | 21.6 | ATR 72-500 | 3.5% | 16.5 | A320-232 | 4.6% | 24.2 |
| 9 | B737-75B | 3.2% | 6.5 | ATR 42-320 | 2.7% | 20.9 | B737-76N | 3.2% | 13.8 | E190-100IGW | 2.7% | 10.8 |
| 10 | B737-3K9 | 3.1% | 17.0 | EMB-120RT | 2.3% | 22.9 | E190-100LR | 3.0% | 5.6 | B737-76N | 2.5% | 17.7 |
| | Top 10 | 59.2% | 14.1 | Top 10 | 68.6% | 13.9 | Top 10 | 64.0% | 11.4 | Top 10 | 80.5% | 12.7 |

*Sources: National Civil Aviation Agency's (ANAC) Brazilian Aeronautical Registry (RAB), Active Scheduled Flight Report (VRA), and Integrated Civil Aviation System (SINTAC); websites planelogger.com, airfleets.net, jetphotos.com, aviacaopaulista.com; all figures computed with authors' calculations. Years since launch measured at the model variant level.*

It can be seen in Table 3 that, among the top 10 most operated model variants, the average age dropped from 14.1 years in 2004 to 12.7 years in 2016 - a 10% drop. In 2004, the three mostly operated model variants were Airbus' A320-232 (12.2 years, measured at the model variant level), Fokker's F100 (17.6 years), and Boeing's B737-76N (5.8 years). In 2016, all three mostly operated variants had lower age, namely, Boeing's B737-8EH (WL) (10.4 years), Embraer's E190-200IGW (9.6 years), and ATR's ATR 72-600 (5.1 years). These figures suggest that the fleet of Brazilian airlines have progressively modernized in the period.

*3.2. Data*

Our primary dataset consists of a panel of 544 domestic city-pairs of the domestic airline industry in Brazil, with monthly observations between January 2004 and December 2018. The full sample contains observations of 9.5 million flights of 38 different aircraft model types and 145 aircraft variants. More than 99 percent of the sample flights were operated with aircraft from 5 manufacturers: Boeing (39.2%), Airbus (36.1%), Embraer (13.9%), ATR (7.3%), and Fokker (3.2%). A minimal market share belongs to LET/Aircraft Industries (0.2%) and CESSNA (0.1%), typically on low-density routes. The participation of manufacturers in the sample has presented notable dynamics over the analyzed decades. In 2004, Airbus' market share was 24.2%, moving to leadership in 2018 with 41.4%. In the same period, Boeing's participation fell from 59.9% (2004) to 29.7% (2018). Embraer had a 5.9% stake in 2004, having dropped to 2.8% in 2007. After Azul Airlines' entry, in 2008, the manufacturer has notably recovered and expanded, reaching around 20.2% at the end of the sample period. The number of aircraft models kept relatively steady in the period, from 22 (2004) to 19 (2018) in operation. However, the number of aircraft types went from 65 (2004) to 40 (2018). Flights operated within the months of the 2014 FIFA World Cup–namely, June and July of that year–are missing in the data set. We restrict our attention to passenger flights and group multiple airports belonging to the same area[14] We discard routes with less than one hundred passengers in a month and with less than six observations in the sample period. In its online database, air transport data in Brazil are publicly available from the National Civil Aviation Agency (ANAC). ANAC supplies data on all the scheduled flights with the Active Scheduled Flight Historical Data Series (VRA). Another ANAC's data set is the Air Transport Statistical Database, with each city-pair/airline's operational flight data by month.[15] We collected jet fuel price information on the National Agency for Petroleum, Natural Gas and Biofuels' (ANP) website. Our sources of fleet characteristics of each airline are ANAC' Brazilian Aeronautical Registry (RAB) and the websites planelogger.com, airfleets.net, jetphotos.com, and aviacaopaulista.com.

### *3.3. Empirical model*

Equation (1) presents our empirical model of technological innovation adoption through fleet modernization of airlines in Brazil.

$$\Delta FLMOD_{k,t} = \beta_1 \Delta FLMOD_{k,t-h} + \beta_2 ASK_{k,t-h} + \beta_3 GROWTH\ SR_{k,t-h} + \beta_4 GROWTH\ LR_{k,t-h}$$
$$+ \beta_5 AGE_{k,t-h} + \beta_6 FUEL\ PRICE_{k,t-h} + \beta_7 FUELPRPK_{k,t-h} + \beta_8 HUB_{k,t-h} \quad (1)$$
$$+ \beta_9 LCC_{k,t-h} + \beta_{10} HHI_{k,t-h} + \beta_{11} HHI2_{k,t-h} + \beta_{12} HHI\ OEM_{k,t-h}$$
$$+ \beta_{13} TREND_t + \beta_{14} TREND_t \times REC_t + u_{k,t},$$

where $k$ denotes the city-pair, $t$ the sample periods ($t = 1, \ldots, 174$ months), and $h$ is a time lag to represent the horizon for airline strategic fleet planning. Using lags reflects the lead-times between the placement of an order by the carrier and the OEMs' delivery. With such a procedure, we aim to capture the ongoing market conditions at the time of capacity decision-making regarding airlines' innovation adoption. We set $h = 24$ because the median delivery lead-time for Boeing and Airbus passenger aircraft was 2.3 years in the 2010s.[16] In the robustness check section, we experiment with other values for $h$ and provide further discussion. Below, we discuss the other components of Equation (1).[17]

---

[14], namely, São Paulo, Rio de Janeiro, and Belo Horizonte metroplexes.

[15] See www.nectar.ita.br/avstats for a detailed description of Brazilian air transport data, with links to the databases.

[16] Source: Global Airline Industry/Commercial Aircraft Backlog: 2000 to 2016 – Airbus, Boeing, and Flightglobal, with own calculations.

[17] To allow the interpretation of regression coefficients as elasticities, we use natural logarithms of all variables, except for fractions and indexes ranging between 0 and 1, rates, and dummies.

- $\Delta FLMOD_{k,t}$ is a proxy for technological innovation adoption through the introduction of new generation aircraft by airlines on city-pair $k$ and time $t$. $FLMOD_{k,t}$ is an indicator of fleet modernization, equal to the logarithm of the inverse of the mean aircraft model type age of the airplanes assigned to flights on the route, multiplied by 100, that is, ln ((1/mean model's age) ×100). The higher the model type's age, the lower the FLMOD indicator. We calculate the ages considering the date of the first flight of that model type's first aircraft. In most cases, the first flight date is the reference. In the absence of this information, we use the date of aircraft delivery. We collect the information on each aircraft registration's first flight and delivery dates from the websites planelogger.com, airfleets.net, jetphotos.com, and aviacaopaulista.com. For example, on July 23, 2018, Gol Airlines operated flight number 1520 between São Paulo (GRU) and Fortaleza (FOR), with the aircraft registration PR-GTP, model type Boeing 738, and model variant 737-8EH(WL). According to the website planelogger.com, the delivery date of that aircraft was August 9, 2007. At the GLO1520 flight, that airplane was almost eleven years old, with Gol itself being the first operator.[18] The 737-8EH(WL) variant had its first unit delivered worldwide by Boeing to the same airline, approximately one year earlier, on August 3, 2006.[19] Thus, that aircraft model variant's age was, on the date of the flight's operation, twelve years. Finally, the model type 737-800 was launched by Boeing in 1994, with planelogger.com identifying its first flight on July 31, 1997.[20] Thus, on the GLO1520 flight operations' date, the airplane's model age was 21 years. When developing the FLMOD variable in our case study, we utilize the aircraft models' mean age on each route and period. In an alternative version of the model, we use the mean age aircraft model variants. Both definitions of FLMOD allow assessing how modern was the aircraft assigned by the airline to the city-pair at the time. For the analysis of innovation adoption in the market, we employ a technological change metric, utilizing the difference $\Delta FLMOD_{k,t} = FLMOD_{k,t} - FLMOD_{k,t-h}$.

- $\Delta FLMOD_{k,t-h}$ is equal to the change in FLMOD observed with $h$ lags, a proxy for the status of technological change at the strategic fleet planning time. Ceteris paribus, an increase in innovation in one period, may inhibit the adoption of innovation by the same airline in a later period, given the aircraft's life cycle. On the other hand, an increase in innovation by technology enthusiasts and early adopters–companies that place greater value on the potential performance-enhancing allowed by novelties in the production process–can induce innovative behavior by pragmatists, conservatives, and laggards–companies that place greater value on solutions and convenience – Rogers (1995), Treloar (1999).

---

[18] See www.planelogger.com/Aircraft/Registration/PR-GTP.
[19] See www.planelogger.com/Aircraft/Registration/PR-GTB/521605.
[20] See www.planelogger.com/Aircraft/Registration/TC-SNY/516455.

- ASK$_{k,t-h}$ is the total available seats-kilometers on the route, divided by 100 (in logarithm); with this variable, we aim to capture the effect of the scale of operations of airlines at the route level, analyzing whether there is any preference for denser city-pairs when introducing modern aircraft.

- GROWTH SR$_{k,t-h}$, and GROWTH LR$_{k,t-h}$ are proxies for market growth in the short and long runs. GROWTH SR$_{k,t-h}$ = ASK$_{k,t-h}$ - ASK$_{k,t-h-12}$, whereas GROWTH LR$_{k,t-h}$ = ASK$_{k,t-h}$ - ASK$_{k,t-h-24}$. The prospect of greater economic growth may motivate expansion plans with greater adoption of more modern aircraft in the strategic fleet planning and management by airlines.

- AGE$_{k,t-h}$ is a proxy for the fleet's mean age assigned to the city-pair (in logarithm). We consider each aircraft's age–that is, each tail number–used on the date of each flight operated in the sample period. As with the variable FLMOD, we use the aircraft's first flight date, according to the information found for each aircraft registration on the websites planelogger.com, airfleets.net, jetphotos.com, and aviacaopaulista.com. An aircraft's age is related to the efficiency losses during its useful life due to the increase in structural problems such as corrosion and fatigue and the weight gain resulting from Maintenance, Repair, and Overhaul (MRO) procedures over the years. According to Duncan (2016), repairs and changes over an aircraft's life are the primary sources of weight changes. However, as an airplane ages, its weight also naturally increases due to debris and dirt collecting in hard-to-reach locations and moisture insulation in the cabin. These gains in weight with age can impair its energy efficiency across time.

- FUEL PRICE$_{k,t-h}$ is a proxy for the jet A1 fuel price in deflated local currency (in logarithm). We add to that metric a proxy for the Brazilian state tax burden on the jet fuel burned on domestic flights, with rates ranging from 3% to 25% depending on the state and period in the sample. FUELP$_{k,t}$ is computed as the minimum average jet fuel price observed at the endpoint cities of a route.

- FUELPRPK$_{k,t-h}$ is equal to the total fuel burn in liters by the scheduled flights on the city-pair, divided by the total revenue-passenger kilometers, and multiplied by 1000 (in logarithm); The higher the fuel intensity of the carrier, the more relevant fleet rollover and fleet modernization strategies at the fleet planning time.

- HUB$_{k,t-h}$ is the maximum percentage of passengers connecting in the city pair's cities of origin and destination. It controls the intensity of the hub operations by carriers. We utilize the US Federal Aviation Administration's definition of a "large hub," and consider only cities containing

- more than one percent of national domestic traffic. In principle, assigning modern aircraft from/to its hubs would be a priority for a network carrier.

- $LCC_{k,t}$ is a dummy of the presence of flights of low-cost carriers (LCC) Gol and Azul airlines when they were new entrants in the market. We consider "new entry" as a period of approximately four years the carrier's startup in the industry. For Gol, we set this dummy variable equal to 1 from January 2001 to May 2005, a period after which its major rivals ceased a codeshare agreement. For Azul, we assigned LCC with one from December 2008 to May 2012, when it announced a merger with regional carrier Trip airlines.

- $HHI_{k,t}$ is the Herfindahl-Hirschman index of city-pair concentration based on the carriers' market shares of revenue passengers. To assess the possible non-linear effects of the market concentration variable, we also use its quadratic term $HHI2_{k,t}$. With $HHI_{k,t}$ and $HHI2_{k,t}$, we aim at testing the hypothesis of the inverted-U relationship between competition and innovation adoption in a framework adapted from Aghion et al. (2005).

- $HHI\ OEM_{k,t}$ is the Herfindahl-Hirschman index of city-pair concentration based on each manufacturer's aircraft flight share–Original Equipment Manufacturer, OEM. With this variable, we aim to control the effects of the competitive dynamics from the rivalry between Boeing and Airbus and, from the mid of the sample period, by the quick spread of Embraer's new airplanes in Brazil through Azul Airlines' acquisitions. The less concentrated the market is concerning OEMs' presence, the greater the competition between them and possibly the greater incentive to purchase more modern aircraft by the airlines. Additionally, the greater the concentration, perhaps the more significant the volume of orders for large manufacturers, opening up the possibility of more significant purchase discounts by scale.

- $TREND_t$ is a time trend variable to control for unobserved technical progress, T = 1, 2, ..., 174;

- $TREND_t \times REC_t$ is an interaction variable designed to control a possible structural break in the trend due to Brazil's mid-2010s recession. $REC_t$ is a dummy variable assigned with one from April 2014 until December 2016–a technical recession period.

- $u_{k,t}$ is the composite error terms of the panels, and the βs are the parameters to be estimated.

Table 4 presents the descriptive statistics and the sources of each of the main variables.[21]

---

[21] See the supplemental material for statistical analysis of the data, with a broad set of diagnostic tests.

**Table 4–Descriptive statistics of the model variables**

| Variable | Description | Metric | Mean | SD. | Min. | Max. | Sources |
|---|---|---|---|---|---|---|---|
| AGE (lagged) | fleet age | years (ln) | 6.60 | 0.60 | 1.88 | 8.37 | (i), (iii), (v) |
| ASK (lagged) | available seat-kilometers | count × km (ln) | 11.03 | 1.77 | 4.04 | 14.83 | (ii) |
| FUEL PRICE (lagged) | price per liter | deflated reais (ln) | 0.93 | 0.22 | 0.16 | 1.43 | (iv), (vi) |
| FUELPRPK (lagged) | fuel burn per revenue passenger-km | 1000 per (count x km) (ln) | 4.05 | 0.45 | 2.89 | 7.68 | (ii), (iv), (vi) |
| GROWTH LR (lagged) | long-run capacity growth | rate | 0.38 | 2.21 | -1.00 | 178.89 | (ii) |
| GROWTH SR (lagged) | short-run capacity growth | rate | 0.26 | 2.50 | -1.00 | 161.40 | (ii) |
| HHI (lagged) | market concentration | index [0,1] | 0.63 | 0.25 | 0.21 | 1.00 | (ii) |
| HHI OEM (lagged) | concentration of aircraft manufacturers | index [0,1] | 0.67 | 0.26 | 0.25 | 1.00 | (i), (ii), (iii) |
| HUB (lagged) | proportion of hub passengers | fraction | 0.16 | 0.09 | 0.00 | 0.47 | (ii) |
| LCC (lagged) | young LCC presence | dummy | 0.12 | 0.33 | 0.00 | 1.00 | (i) |
| ΔFLMOD (model) | fleet modernization (model) change | index (ln), difference | 0.00 | 0.40 | -3.08 | 3.89 | (i), (ii), (iii), (v) |
| ΔFLMOD (model) (lagged) | fleet modernization (model) change | index (ln), difference | 0.02 | 0.38 | -3.03 | 3.89 | (i), (ii), (iii), (v) |
| ΔFLMOD (variant) | fleet modernization (variant) change | index (ln), difference | 0.02 | 0.42 | -3.08 | 3.85 | (i), (ii), (iii), (v) |

*Sources: (i) Active Scheduled Flight Historical Data Series—VRA; (ii) Air Transport Statistical Database; (iii) Brazilian Aeronautical Registry—RAB; (iv) National Agency for Petroleum, Natural Gas and Biofuels—ANP; (v) websites planelogger.com, airfleets.net, jetphotos.com, and aviacaopaulista.com; (vi) state-specific legislation and online media news; all figures computed with authors' calculations. See the supplemental material for a correlation table and other statistical analyses of the variables.*

Figure 1 shows, in a scatter diagram, the mean rate of change in fleet modernization (ΔFLMOD) against the mean market concentration (HHI) over the years in the sample period. In this diagram, two concepts of fleet modernization are displayed: aircraft model type and aircraft model variant. The dispersion of points contains some evidence suggesting an inverted-U shaped relationship between the variables. The figure roughly indicates the probability of observing a higher ΔFLMOD in routes with market concentration levels between 0.45 and 0.55. This rudimentary evidence motivates our proposed econometric modeling and hypothesis testing regarding that association.

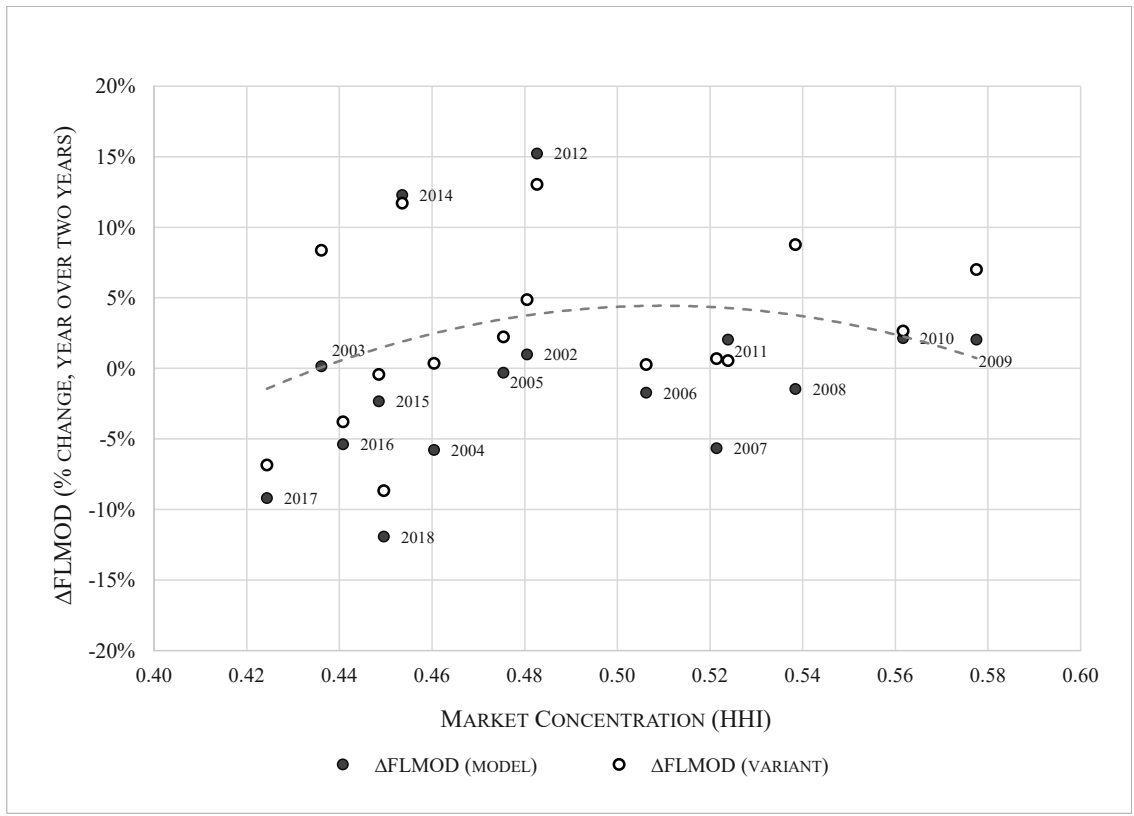

**Figure 1–Innovation adoption (ΔFLMOD) against market concentration (HHI)**

## 3.4. Estimation strategy

We treat fleet modernization by airlines as a complex decision-making process related to several technological, operational, and competitive influences. These factors are not fully observed by the econometrician. Our empirical approach to Equation (1) aims to deal with such unobservability issues. We perform the estimation utilizing the econometric method of high dimensional sparse (HDS) regression models of Belloni et al. (2012), Belloni, Chernozhukov, & Hansen (2014a,b), and Chernozhukov, Hansen, & Spindler (2015)–a Post-Double Selection Estimator, PDS-LASSO. This procedure allows flexibility in allowing a broad set of regressors and nuisance parameters to control for unobserved effects in the estimation that could produce omitted variables bias. In parallel, the utilization of the least absolute shrinkage and selection operator (LASSO) of Tibshirani (1996) inhibits model overfitting by regressor penalization.

We utilize the following controls as nuisance parameters in our model: (i) route fixed effects, aiming to control the city-pair-specific and endpoints-specific, time-invariant idiosyncrasies that may be correlated with fleet assignment–route distance, and city locational factors, among others–with this procedure, we employ an FE-PDS-LASSO estimating approach; (ii) dummy variables to account for route-specific quarter-to-quarter seasonality; and (iii) dummy variables to account for the presence of each airline on the region-pair of the respective city-pair, also in a quarter-to-quarter basis; these controls aim to proxy for the airline-specific, seasonal changes in fleet assignment across markets. Our model contains up to 4,079 controls of type (ii) and 1,471 controls of type (iii), all subject to LASSO inactivation through its parameter penalization procedure. Also, except for the ΔFLMOD, ASK, GROWTH SR, GROWTH LR, and AGE variables, we include all other covariates in the models in the set of variables penalized by the LASSO procedure.

Concerning diagnosis tests, we develop a broad set of statistical assessments of the data set and our empirical approach's potential problems. In particular, we perform analyses of correlations and multicollinearity, heteroscedasticity, autocorrelation, normality, and model misspecification. Additionally, we performed tests of types of panel model, unit root, and cointegration. In particular, our tests confirm the presence of both heteroscedasticity and autocorrelation, which lead us to utilize the Newey-West procedure to adjust the standard error estimates of the parameters. In the implementation of the estimating approach, we use the procedures of Ahrens, Hansen & Schaffer (2020) to estimate cluster-robust penalty loads to tackle heteroscedasticity issues in regression further. In this setup, each city-pair is a different cluster. Concerning multicollinearity in the estimation, we extract mean and maximum VIF statistics of, respectively, 11.97 and 74.02. These results confirm the presence of relevant multicollinearity problems and make us cautious when interpreting the nonsignificant variables' results in our approach. All diagnosis tests and analyses can be found in the supplemental material.

To challenge our baseline model using FE-PDS-LASSO, we also conduct alternative estimation approaches. For such an objective, we utilize fixed-effects versions of the Bayesian model averaging (BMA) of Leamer (1978), and the Weighted-average least squares (WALS) of Magnus, Powell, and Prüfer (2010). As discussed in De Luca & Magnus (2011), these model-averaging estimators allow to taking into account the uncertainty inherent to the estimation and the model selection processes when making inference on the parameters of interest. We denote these estimators as "FE-BMA" and "FE-WALS" in our estimation results tables. We also consider a simple version of the models, with the implementation of a pure fixed effects model, denoted as "FE." Finally, we consider a version of the baseline model applied to a reduced sample created by dropping observations related to thin-density routes, defined as those connecting only non-state-capital cities. We denote this sample as "Drop Regional." This experiment is potentially more appropriate for fleet modernization analysis, as thin-density routes are commonly operated by small regional carriers with turboprop aircraft of low seating capacity.

## 4. Estimation results

Table 5 presents the estimation results of the proposed model of fleet modernization in Brazil. Each column of Table 5 represents a different type of estimator, controls, or sample. It is possible to assess the different specifications' relative performance using the AIC, BIC, RMSE, 4-Fold cross-validation RMSE ("RMSE CV") statistics. According to each criterion, a number between square brackets next to these statistics indicates the statistic value's rank across columns.

Column (8) contains our preferred specification results – namely, the FE-PD-LASSO model with city-pair-quarter and region-pair-quarter-airline controls, and the sample with thin-density, regional markets dropped. Most statistics displayed at the table's bottom are the first or second among the columns in this specification. Regardless of these criteria, we note that the result of most variables is robust across specifications. More specifically, the variables ΔFLMOD, ASK, GROWTH LR, AGE, FUEL PRICE, FUELPRPK, HHI OEM, and TREND × REC keep the same result concerning sign and statistical significance in all columns.

Column (1) displays one basic version, not considering the HH2 variable, which we introduce in Column (2). From Column (3), we also include the variable HHI OEM as a way to challenge this possible non-linear association between market concentration and fleet modernization. We note that, in principle, the variable HHI is not statistically significant when inserted separately in Column (1). Once we insert the quadratic term HHI2, both variables become statistically relevant, providing some evidence supporting the U-inverted relationship hypothesis, as in Aghion et al. (2005). This result is not altered when we insert the variable HHI OEM–Column (2)–, which aims to control a possible confounding effect of manufacturers' concentration in the relationship between HHI and ΔFLMOD. The inverted-U innovation adoption pattern is also not rejected in the other specifications in Table 5, in Columns (3)–(9).

The other results of Table 5 are as follows. Firstly, the estimated coefficient of ΔFLMOD is negative, indicating the existence of alternating innovation adoption cycles by airlines, in which periods of more intense acquisition of new generation aircraft are followed by periods of less intense purchase. This behavior suggests that the opportunity cost of adopting new technologies from the manufacturer increases as the fleet modernizes, indicating that, ceteris paribus to market growth, the investment made more recently in aircraft acquisitions takes some time to mature before engaging in new additions. Additionally, from a purely financial point of view, the level of airlines' indebtedness increases with the acquisition of new aircraft, which probably inhibits new purchases in the subsequent period. Finally, the innovation adoption cycles could also be explained by the fact that the launch of new models and variants in this industry is not so frequent, but actually much slower than in other markets, such as automotive, for example.

The ASK variable was not statistically significant, which indicates that routes with greater installed seating capacity are not necessarily operated with more modern aircraft. However, the long-term growth–GROWTH LR–in this capacity proved to be statistically significant and positively related to ΔFLMOD. As expected, the AGE variable had a positive estimated coefficient, indicating that airlines seek to introduce new generation aircraft when they implement their fleet rollover to avoid fleet aging policies. Also, the estimated positive relationship between ΔFLMOD and FUEL PRICE in all model specifications provides evidence of an incentive for airlines to engage in fleet modernization, in the long run, to save operational costs with fuel. The same reasoning applies to the result of the FUELPRPK variable.

Regarding the HUB and LCC variables, our results suggest positive effects of these variables. However, we consider it to be limited evidence, given that these variables suffer from instability in the specifications in which deep controls were used with the use of nuisance parameters, from Columns (6)–(9). For example, the LCC variable is inactivated by the PDS-LASSO in Columns (7) and (8).

Concerning the trend variables (TREND and TREND × REC), we note the following results. First, we find some evidence that the intensity of fleet modernization seems to increase over the years, as estimated by the variable TREND, possibly due to the unobserved technical progress in aircraft engine, design, and aerodynamics, among other factors. However, this variable is not statistically significant in Column (6), suggesting that it may be correlated with the city-pair and region-pair-airline-specific quarter controls. Additionally, we find evidence that the fleet modernization's growth trend is amplified during the recession period of 2014-2016, as assessed by the estimated coefficients of TREND × REC.

For the performance of the FE-PDS-LASSO estimator in selecting the high-dimension controls, Table 5 allows some considerations. The effect of model shrinkage can be seen in the active control counts displayed at the bottom of that table in Columns (7)–(9). From the 4,079 city-pair-quarter-controls, the LASSO procedure selected between 454 and 461. Also, of the 1,471 region-pairs-quarter-airline controls, LASSO selected between 58 and 60. The percentage of controls' activation was approximately 11% and 4%, respectively, illustrating the estimator's effectiveness in setting the model's nuisance parameters.

Table 5–Estimation results: fleet modernization change (ΔFLMOD)–aircraft level: model

| | (1) ΔFLMOD | (2) ΔFLMOD | (3) ΔFLMOD | (4) ΔFLMOD | (5) ΔFLMOD | (6) ΔFLMOD | (7) ΔFLMOD | (8) ΔFLMOD | (9) ΔFLMOD |
|---|---|---|---|---|---|---|---|---|---|
| ΔFLMOD (lagged) | -0.1836*** | -0.1791*** | -0.1791*** | -0.1786*** | -0.1802*** | -0.2193*** | -0.1828*** | -0.1685*** | -0.1861*** |
| ASK (lagged) | -0.0091 | -0.0115 | -0.0046 | -0.0000 | -0.0035 | 0.0088 | 0.0019 | -0.0013 | 0.0009 |
| GROWTH SR (lagged) | 0.0021* | 0.0021* | 0.0021* | 0.0015 | 0.0017*** | 0.0018* | 0.0023* | 0.0022* | 0.0023* |
| GROWTH LR (lagged) | 0.0058*** | 0.0057*** | 0.0058*** | 0.0057*** | 0.0053*** | 0.0040*** | 0.0056*** | 0.0054*** | 0.0055*** |
| AGE (lagged) | 0.2917*** | 0.2946*** | 0.2957*** | 0.2961*** | 0.2916*** | 0.3046*** | 0.2857*** | 0.3001*** | 0.2831*** |
| FUEL PRICE (lagged) | 0.1669*** | 0.1506*** | 0.1515*** | 0.1516*** | 0.1527*** | 0.1318*** | 0.1396*** | 0.1560*** | 0.1402*** |
| FUELPRPK (lagged) | 0.0290** | 0.0342*** | 0.0380*** | 0.0398*** | 0.0398*** | 0.0460*** | 0.0356*** | 0.0384*** | 0.0360*** |
| HUB (lagged) | 0.1091** | 0.1418*** | 0.1500*** | 0.1469*** | 0.1467*** | 0.0180 | 0.0924* | 0.1633*** | – |
| LCC (lagged) | 0.0953*** | 0.1092*** | 0.1064*** | 0.1062*** | 0.1059*** | 0.0644*** | – | – | <check> |
| HHI (lagged) | -0.0178 | 0.9459*** | 0.9546*** | 0.9565*** | 0.9497*** | 0.4161*** | 0.6950*** | 0.7241*** | 0.7057*** |
| HHI2 (lagged) | | -0.6982*** | -0.7232*** | -0.7226*** | -0.7211*** | -0.3188*** | -0.5533*** | -0.5795*** | -0.5556*** |
| HHI OEM (lagged) | | | 0.0673*** | 0.0710*** | 0.0673*** | 0.0580*** | 0.0822*** | 0.0768*** | 0.0763*** |
| TREND | 0.0176*** | 0.0184*** | 0.0188*** | 0.0186*** | 0.0186*** | 0.0001 | 0.0143*** | 0.0136*** | 0.0146*** |
| TREND × REC | 0.0023*** | 0.0030*** | 0.0029*** | 0.0029*** | 0.0028*** | 0.0046*** | 0.0033*** | 0.0031*** | 0.0034*** |
| Sample | Full | Full | Full | Full | Full | Full | Full | Drop Regional | Full |
| Estimator | FE | FE | FE | FE-BMA | FE-WALS | FE | FE-PDS-LASSO | FE-PDS-LASSO | FE-PDS-LASSO |
| City-Pair-Qtr controls | No | No | No | No | No | 4079/4079 | 513/4079 | 487/4079 | 529/4079 |
| Reg-Pair-Qtr-Airl controls | No | No | No | No | No | 1471/1471 | 74/1471 | 86/1471 | 74/1471 |
| Adj R2 Statistic | 0.2265 [9] | 0.2307 [6] | 0.2313 [5] | 0.2390 [2] | 0.2390 [2] | 0.2337 [4] | 0.2290 [8] | 0.2430 [1] | 0.2302 [7] |
| AIC Statistic | 33,063 [7] | 32,765 [6] | 32,727 [3] | 32,733 [4] | 32,733 [4] | 22,144 [1] | 33,468 [9] | 28,936 [2] | 33,399 [8] |
| BIC Statistic | 33,170 [6] | 32,881 [5] | 32,852 [2] | 32,876 [3] | 32,876 [3] | 22,269 [1] | 38,810 [8] | 34,121 [7] | 38,874 [9] |
| RMSE Statistic | 0.3296 [9] | 0.3287 [8] | 0.3286 [7] | 0.3269 [3] | 0.3269 [3] | 0.2981 [1] | 0.3273 [6] | 0.3172 [2] | 0.3270 [5] |
| RMSE CV Statistic | 0.3728 [4] | 0.3729 [5] | 0.3730 [6] | 0.3671 [2] | 0.3671 [2] | 0.4166 [9] | 0.3760 [8] | 0.3620 [1] | 0.3754 [7] |
| Nr Obervations | 54,338 | 54,338 | 54,338 | 54,338 | 54,338 | 54,338 | 54,338 | 52,177 | 54,338 |

Notes: Columns (1)-(3) and (6) estimated with a fixed-effects procedure; Columns (4) and (5) estimated with, respectively, the Bayesian model averaging (BMA) of Leamer (1978), and the Weighted-average least squares (WALS) of Magnus, Powell, and Prüfer (2010); Columns (7)-(9) estimated with the post-double-selection (PDS) LASSO-based methodology of Belloni et al. (2012, 2014a,b). LASSO penalty loadings account for city-pairs clustering. Post-LASSO estimation is performed with a fixed-effects procedure with standard errors robust to heteroskedasticity and autocorrelation. Control variables estimates omitted; "<check>" indicates the intentional drop of the variable to check the model's robustness; blank cells indicate that the variable was not used; "–" means that the LASSO procedure inactivated the variable; a number between square brackets denotes the rank of the statistic value according to each criterion–from the "best" to the "worse" across columns; ΔFLMOD, ASK, GROWTH SR, GROWTH LR, and AGE not penalized by LASSO; p-value representations: ***$p<0.01$, ** $p<0.05$, * $p<0.10$.

Table 6–Estimation results: fleet modernization change (ΔFLMOD)–aircraft level: variant

| | (1) ΔFLMOD | (2) ΔFLMOD | (3) ΔFLMOD | (4) ΔFLMOD | (5) ΔFLMOD | (6) ΔFLMOD | (7) ΔFLMOD | (8) ΔFLMOD | (9) ΔFLMOD |
|---|---|---|---|---|---|---|---|---|---|
| ΔFLMOD (lagged) | -0.1558*** | -0.1535*** | -0.1536*** | -0.1532*** | -0.1548*** | -0.1769*** | -0.1541*** | -0.1378*** | -0.1546*** |
| ASK (lagged) | -0.0052 | -0.0069 | -0.0050 | -0.0000 | -0.0035 | 0.0053 | -0.0004 | -0.0034 | -0.0012 |
| GROWTH SR (lagged) | 0.0017 | 0.0018 | 0.0018 | 0.0004 | 0.0013** | 0.0012 | 0.0020 | 0.0018* | 0.0020 |
| GROWTH LR (lagged) | 0.0046*** | 0.0045*** | 0.0046*** | 0.0046*** | 0.0041*** | 0.0033*** | 0.0044*** | 0.0043*** | 0.0044*** |
| AGE (lagged) | 0.3477*** | 0.3495*** | 0.3498*** | 0.3502*** | 0.3457*** | 0.3662*** | 0.3436*** | 0.3638*** | 0.3416*** |
| FUEL PRICE (lagged) | 0.1680*** | 0.1571*** | 0.1574*** | 0.1578*** | 0.1586*** | 0.1273*** | 0.1461*** | 0.1660*** | 0.1421*** |
| FUELPRPK (lagged) | -0.0056 | -0.0022 | -0.0011 | 0.0008 | 0.0009 | 0.0042 | -0.0004 | -0.0003 | 0.0012 |
| HUB (lagged) | 0.0903 | 0.1129* | 0.1151** | 0.1113* | 0.1113*** | 0.0198 | 0.0981* | 0.1836*** | – |
| LCC (lagged) | 0.0550*** | 0.0643*** | 0.0635*** | 0.0633*** | 0.0630*** | 0.0174** | – | – | <check> |
| HHI (lagged) | 0.0144 | 0.6572*** | 0.6595*** | 0.6606*** | 0.6550*** | 0.2294*** | 0.4871*** | 0.5350*** | 0.4810*** |
| HHI2 (lagged) | | -0.4658*** | -0.4728*** | -0.4719*** | -0.4707*** | -0.1478** | -0.3649*** | -0.3984*** | -0.3639*** |
| HHI OEM (lagged) | | | 0.0189 | 0.0226** | 0.0193* | 0.0245 | 0.0313* | 0.0203 | 0.0323* |
| TREND | 0.0151*** | 0.0156*** | 0.0157*** | 0.0156*** | 0.0155*** | -0.0001 | 0.0127*** | 0.0125*** | 0.0138*** |
| TREND × REC | 0.0033*** | 0.0038*** | 0.0037*** | 0.0037*** | 0.0037*** | 0.0045*** | 0.0042*** | 0.0040*** | 0.0044*** |
| Sample | Full | Full | Full | Full | Full | Full | Full | Drop Regional | Full |
| Estimator | FE | FE | FE | BMA | WALS | FE | PDS/LASSO | PDS/LASSO | PDS/LASSO |
| City-Pair-Qtr controls | No | No | No | No | No | 4079/4079 | 461/4079 | 454/4079 | 456/4079 |
| Reg-Pair-Qtr-Airl controls | No | No | No | No | No | 1471/1471 | 60/1471 | 69/1471 | 58/1471 |
| Adj R2 Statistic | 0.2543 [7] | 0.2560 [4] | 0.2560 [4] | 0.2633 [3] | 0.2634 [2] | 0.2503 [9] | 0.2548 [6] | 0.2717 [1] | 0.2538 [8] |
| AIC Statistic | 36,571 [7] | 36,447 [4] | 36,446 [3] | 36,457 [6] | 36,453 [5] | 27,188 [1] | 37,042 [8] | 32,586 [2] | 37,110 [9] |
| BIC Statistic | 36,677 [6] | 36,563 [2] | 36,571 [3] | 36,600 [5] | 36,595 [4] | 27,312 [1] | 41,796 [9] | 37,336 [7] | 41,792 [8] |
| RMSE Statistic | 0.3404 [9] | 0.3400 [7] | 0.3400 [7] | 0.3383 [3] | 0.3383 [3] | 0.3122 [1] | 0.3386 [5] | 0.3288 [2] | 0.3389 [6] |
| RMSE CV Statistic | 0.3794 [2] | 0.3796 [3] | 0.3796 [3] | 0.3873 [7] | 0.3873 [7] | 0.4362 [9] | 0.3813 [5] | 0.3683 [1] | 0.3814 [6] |
| Nr Obervations | 54,338 | 54,338 | 54,338 | 54,338 | 54,338 | 54,338 | 54,338 | 52,177 | 54,338 |

Notes: Columns (1)-(3) and (6) estimated with a fixed-effects procedure; Columns (4) and (5) estimated with, respectively, the Bayesian model averaging (BMA) of Leamer (1978), and the Weighted-average least squares (WALS) of Magnus, Powell, and Prüfer (2010); Columns (7)-(9) estimated with the post-double-selection (PDS) LASSO-based methodology of Belloni et al. (2012, 2014a,b). LASSO penalty loadings account for city-pairs clustering. Post-LASSO estimation is performed with a fixed-effects procedure with standard errors robust to heteroskedasticity and autocorrelation. Control variables estimates omitted; "<check>" indicates the intentional drop of the variable to check the model's robustness; blank cells indicate that the variable was not used; "–" means that the LASSO procedure inactivated the variable; a number between square brackets denotes the rank of the statistic value according to each criterion–from the "best" to the "worse" across columns; ΔFLMOD, ASK, GROWTH SR, GROWTH LR, and AGE not penalized by LASSO; p-value representations: ***$p<0.01$, ** $p<0.05$, * $p<0.10$.

Table 6 presents the estimation results considering the variable ΔFLMOD measured in an alternative way, this time using the mean age of the aircraft model variants instead of aircraft model types. Apart from that, the specifications of the columns in Table 6 are identical to those in Table 5. The vast majority of the results obtained in Table 5 remained the same in Table 6. Only the variables FUELPRPK and HHI OEM presented changes. In both cases, the variables were positive and statistically significant in Table 5 and lost their significance in Table 6.

On the other hand, the variables FUEL PRICE, HHI, and HHI2 remained statistically significant. These results, taken together, suggest that increases in fuel prices and, to a certain extent, competition serve as a motivator for introducing new types of aircraft models and new variants of them. However, the introduction of new models is also dictated by the greater or lesser energy efficiency observed in each route–measured by FUELPRPK–, a phenomenon not observed with the introduction of new model variants. It seems that airlines assign new model variants on their routes regardless of whether they need to increase their eco-efficiency specifically in these markets. They apparently only do so by introducing new aircraft models, which may indicate that this seems to be the dimension where the efficiency gains from the adoption of technological innovation are greatest. Therefore, the estimation results suggest that the most impactful incremental technological innovation sources from introducing new commercial aviation equipment are related to the new types of aircraft models of the OEMs and not the new variants launched.

*4.1 Robustness Checks*

In our main specification, to construct the model's regressand and regressors, we use lags. We consider a lag of two years to account for the time horizon between the airline's fleet planning and fleet modernization's actual materialization in the market–the new aircraft's assignment to routes. As discussed before, we base our choice of lag length on the manufacturers' median lead-time between placing an aircraft order and delivery, commonly close to two years in the 2010s. However, when analyzing orders placed over the 2000s, we observe that the median delivery lead-time was 43% higher, around 3.3 years.[22] On the other hand, lead-times have been steadily declining since the early 2010s. Clark (2007) suggests that lead-times may be as little as nine months for single-aisle aircraft without customization requests. He also observes that through leasing contracts, carriers may also "*control the speed at which capacity comes into the market and, crucially, in providing capacity at relatively short notice*" (p. 13). According to the author, operating lessors may offer aircraft for delivery as quickly as three or four months.

---

[22] Source: Global Airline Industry/Commercial Aircraft Backlog: 2000 to 2016 – Airbus, Boeing, and Flightglobal, with own calculations.

To inspect the robustness of our empirical results concerning the lag length setup, we develop a series of checks. We perform the estimates of the same preferred model of Table 5–i.e., Column (8)–, but with changes in the lag used to construct the model variables. We experiment with lags of 1, 1.5, 2.5, 3, 3.5, 4, 4.5, and 5 years. The results of these experiments–displayed in Appendix–, confirmed the robustness of the vast majority of variables used in the main specification. However, about the price of fuel (FUEL PRICE) and market concentration (HHI and HHI2), it was possible to observe some thought-provoking results. For higher lags, above 2.5 years for the case of FUEL PRICE, and above 3 years for HHI and HHI2, the results are no longer significant, and in some cases, there is inactivation by the LASSO procedure. These results indicate that the incentives stemming from market competition and fuel price have a relevant influence on carriers' modernization actions only up to a certain time horizon, after which the effects dissipate. These results are intuitive, given that the airline industry is well-known for high uncertainty levels, associated with challenging costs of capital. In such a market environment, not all carriers may be willing to risk-taking too much in advance in their fleet planning horizons.

## 5. Conclusion

This present paper proposed an econometric model of fleet modernization to inspect airlines' innovation adoption behavior regarding the acquisition of new generation aircraft. We develop a new fleet modernization index considering the launch date of the aircraft model types and variants that operated flights in the Brazilian domestic airline market from 2000 to 2018. We formally tested the existence of an inverted-U shaped relationship between market concentration and fleet modernization, with results supporting that hypothesis in most cases. However, we find that the relationship vanishes when we consider longer fleet planning time horizons. We also find evidence of fuel price signals providing incentives for fleet modernization after up to two years–a result consistent with OEM's lead-times of aircraft deliveries during the sample period. These findings suggest that energy cost rises may provoke boosts in fleet modernization in the long term, with carriers possibly targeting more eco-efficient operations up to two years after an increase in fuel price.

Additionally, our estimation results show only limited evidence for the long-term effects of low-cost carriers and hubs on airlines' strategic fleet planning regarding fleet modernization. Furthermore, the results suggest that the most environmentally friendly incremental technological innovation source from introducing new commercial aviation equipment is related to the new types of aircraft models, not the new variants launched by the OEMs.

Our model has the limitation of strictly focusing on the estimation of an empirical model. We recommend that future studies develop theoretical models on the rationale and incentives for the technological innovation adoption of carriers, as a way to support the econometric approach, aiming at formally incorporating the concepts of Schumpeterian and escape-competition effects of Aghion et al. (2005) to the analysis. This theoretical approach could map the differences between an inter-industry study like the authors', and an intra-industry approach, as with the present study, incorporating these differences in a single conceptual framework.

The study of the behavior of technological innovation adoption by airlines is essential to pinpoint the causes of the continuing increase in the air transport environmental footprint. Through a greater understanding of the carriers' strategic fleet planning incentives for the acquisition of new manufacturers' products, public and corporate policies that pursue greater eco-efficiency in this industry can be improved, in compliance with the internationally stipulated emissions targets.

# References

Aghion, P., Bloom, N., Blundell, R., Griffith, R., & Howitt, P. (2005). Competition and innovation: An inverted-U relationship. The quarterly journal of economics, 120(2), 701-728.

Aghion, P., Blundell, R., Griffith, R., Howitt, P., & Prantl, S. (2009). The effects of entry on incumbent innovation and productivity. The Review of Economics and Statistics, 91(1), 20–32.

Ahrens, A., Hansen, C. B., & Schaffer, M. E. (2020). LASSOPACK: Model selection and prediction with regularized regression in Stata. The Stata Journal, 20(1), 176-235.

Albers, S., Daft, J., Stabenow, S., & Rundshagen, V. (2020). The long-haul low-cost airline business model: A disruptive innovation perspective. Journal of Air Transport Management, 89, 101878.

Askenazy, P., Cahn, C., & Irac, D. (2013). Competition, R&D, and the cost of innovation: Evidence for France. Oxford Economic Papers, 65(2), 293–311.

Baldwin, W., & Scott, J. (2013). Market structure and technological change (Vol. 1). Taylor & Francis.

Belloni, A., Chen, D., Chernozhukov, V., & Hansen, C. (2012). Sparse models and methods for optimal instruments with an application to eminent domain. Econometrica, 80(6), 2369-2429.

Belloni, A., Chernozhukov, V., & Hansen, C. (2014a). Inference on treatment effects after selection among high-dimensional controls. The Review of Economic Studies, 81(2), 608-650.

Belloni, A., Chernozhukov, V., & Hansen, C. (2014b). High-dimensional methods and inference on structural and treatment effects. Journal of Economic Perspectives, 28(2), 29-50.

Beneito, P., Rochina-Barrachina, M. E., & Sanchis, A. (2017). Competition and innovation with selective exit: An inverted-U shape relationship? Oxford Economic Papers, 69(4), 1032–1053.

Bérubé, C., Duhamel, M., & Ershov, D. (2012). Market incentives for business innovation: Results from Canada. Journal of Industry, Competition and Trade, 12(1), 47-65.

Brueckner, J. K., & Pai, V. (2009). Technological innovation in the airline industry: The impact of regional jets. International Journal of Industrial Organization, 27(1), 110-120.

Caves, D. W., Christensen, L. R., & Tretheway, M. W. (1984). Economies of density versus economies of scale: why trunk and local service airline costs differ. The RAND Journal of Economics, 471-489.

Chandler Jr., A. D. (1994). The competitive performance of US industrial enterprises since the Second World War. The Business History Review, 1-72.

Chernozhukov, V., Hansen, C., & Spindler, M. (2015). Post-selection and post-regularization inference in linear models with many controls and instruments. American Economic Review, 105(5), 486-90.

Clark, P. (2017). Buying the big jets: fleet planning for airlines. Taylor & Francis.

Damanpour, F. (1991). Organizational Innovation: A Meta-Analysis of Effects of Determinants and Moderators. Academy of Management Journal, 34(3), 555–590.

Duncan (2016) Weight and Balance Handbook. Oklahoma City: US Department of Transportation - Federal Aviation Administration, Airman Testing Standards Branch, FAA-H-8083-1B.

Frambach, R. T., & Schillewaert, N. (2002). Organizational innovation adoption: A multi-level framework of determinants and opportunities for future research. Journal of Business Research, 55(2), 163–176.

Franke, M. (2007). Innovation: The winning formula to regain profitability in aviation?. Journal of air transport management, 13(1), 23-30.

Fudenberg, D., & Tirole, J. (1984). The fat-cat effect, the puppy-dog ploy, and the lean and hungry look. The American Economic Review, 74(2), 361-366.

Holloway, S. (2008). Straight and level: practical airline economics. Ashgate Publishing, Ltd.

Jansen, P. W., & Perez, R. E. (2016). Coupled optimization of aircraft families and fleet allocation for multiple markets. Journal of Aircraft, 53(5), 1485-1504.

Kamien, M. I., & Schwartz, N. L. (1982). Market structure and innovation. Cambridge University Press.

Kilpi, J. (2007). Fleet composition of commercial jet aircraft 1952–2005: Developments in uniformity and scale. Journal of Air Transport Management, 13(2), 81-89.

Kreps, D. M., & Wilson, R. (1982). Reputation and imperfect information. Journal of Economic Theory, 27(2), 253-279.

Leamer, E. E. (1978). Specification Search: Ad Hoc Inference with Nonexperimental Data. New York: Wiley.

Magnus, J. R., Powell, O., & Prüfer, P. (2010). A comparison of two model averaging techniques with an application to growth empirics. Journal of Econometrics, 154(2), 139-153.

Milgrom, P., & Roberts, J. (1982). Predation, reputation, and entry deterrence. Journal of Economic Theory, 27(2), 280-312.

Narcizo, R. R., Oliveira, A. V. M., & Dresner, M. E. (2020). An empirical model of airline fleet standardization in Brazil: Assessing the dynamic impacts of mergers with an events study. Transport Policy, 97, 149-160.

Nash, C., Matthews, B., & Smith, A. (2020). The impact of rail industry restructuring on incentives to adopt innovation: A case study of Britain. Proceedings of the Institution of Mechanical Engineers, Part F: Journal of Rail and Rapid Transit, 234(3), 331-337.

Peneder, M. (2012). Competition and Innovation: Revisiting the inverted U-relationship. J Ind Compet Trade, 12, 1–5.

Peeters, P., Higham, J., Kutzner, D., Cohen, S., & Gössling, S. (2016). Are technology myths stalling aviation climate policy?. Transportation Research Part D: Transport and Environment, 44, 30-42.

Polder, M., & Veldhuizen, E. (2012). Innovation and competition in the Netherlands: Testing the inverted-U for industries and firms. Journal of Industry, Competition and Trade, 12(1), 67-91.

Rogers, E. M. (1995). Diffusion of innovations (4th edition). The Free Press.

Tibshirani, R. (1996). Regression shrinkage and selection via the Lasso. Journal of the Royal Statistical Society: Series B (Methodological), 58(1), 267-288.

Treloar, A. E. (1999). Products and Processes: How Innovation and Product Life-Cycles Can Help Predict the Future of the Electronic Scholarly Journal. In ELPUB.

Zou, L., Yu, C., & Dresner, M. (2015). Fleet standardisation and airline performance. Journal of Transport Economics and Policy (JTEP), 49(1), 149-166.

**Appendix Table A1–Robustness checks estimation results: fleet modernization change (ΔFLMOD)–aircraft level: model**

|  | (1) $\Delta_{1y}$FLMOD | (2) $\Delta_{1.5y}$FLMOD | (3) $\Delta_{2y}$FLMOD | (4) $\Delta_{2.5y}$FLMOD | (5) $\Delta_{3y}$FLMOD | (6) $\Delta_{3.5y}$FLMOD | (7) $\Delta_{4y}$FLMOD | (8) $\Delta_{4.5y}$FLMOD | (9) $\Delta_{5y}$FLMOD |
|---|---|---|---|---|---|---|---|---|---|
| ΔFLMOD (lagged) | -0.1472*** | -0.1311*** | -0.1685*** | -0.2151*** | -0.3129*** | -0.3660*** | -0.3586*** | -0.3672*** | -0.4209*** |
| ASK (lagged) | 0.0191*** | 0.0213*** | -0.0013 | -0.0166* | -0.0417*** | -0.0517*** | -0.0693*** | -0.0804*** | -0.0779*** |
| GROWTH SR (lagged) | 0.0022** | 0.0017** | 0.0022* | 0.0015 | 0.0004 | 0.0014 | 0.0005 | 0.0017 | 0.0002 |
| GROWTH LR (lagged) | 0.0002 | 0.0008** | 0.0054*** | 0.0029*** | 0.0013** | 0.0021*** | 0.0014** | 0.0010* | 0.0013** |
| AGE (lagged) | 0.1926*** | 0.2498*** | 0.3001*** | 0.3303*** | 0.3130*** | 0.2932*** | 0.2923*** | 0.2868*** | 0.2421*** |
| FUEL PRICE (lagged) | 0.1238*** | 0.1502*** | 0.1560*** | – | – | – | – | – | – |
| FUEL/RPK (lagged) | 0.0058 | 0.0304** | 0.0384*** | 0.0328** | 0.0436*** | 0.0208 | 0.0502** | 0.0587*** | 0.0838*** |
| LCC (lagged) | – | – | – | – | – | – | – | -0.0476*** | -0.0753*** |
| HUB (lagged) | 0.1578*** | 0.1269*** | 0.1633*** | 0.1858*** | 0.2806*** | 0.3450*** | – | 0.1360* | -0.0398 |
| HHI (lagged) | 0.3391*** | 0.5867*** | 0.7241*** | 0.9143*** | -0.0500* | -0.0290 | -0.0264 | -0.0054 | -0.0540* |
| HHI2 (lagged) | -0.2582*** | -0.4657*** | -0.5795*** | -0.7377*** | – | – | – | – | – |
| HHI OEM (lagged) | 0.0174 | 0.0551*** | 0.0768*** | 0.0969*** | 0.0763*** | 0.1118*** | 0.1030*** | 0.0803*** | 0.0730*** |
| TREND | 0.0064*** | 0.0108*** | 0.0136*** | 0.0117*** | 0.0119*** | 0.0121*** | 0.0198*** | 0.0156*** | 0.0153*** |
| TREND × REC | 0.0009* | 0.0017*** | 0.0031*** | 0.0098*** | 0.0109*** | 0.0137*** | 0.0161*** | 0.0197*** | 0.0230*** |
| Sample | Drop Regional | Drop Regional | Drop Regional | Drop Regional | Drop Regional | Drop Regional | Drop Regional | Drop Regional | Drop Regional |
| Estimator | FE/PDS/LASSO | FE/PDS/LASSO | FE/PDS/LASSO | FE/PDS/LASSO | FE/PDS/LASSO | FE/PDS/LASSO | FE/PDS/LASSO | FE/PDS/LASSO | FE/PDS/LASSO |
| City-Pair-Qtr controls | 481/4079 | 401/4079 | 487/4079 | 433/4079 | 364/4079 | 388/4079 | 357/4079 | 220/4079 | 261/4079 |
| Reg-Pair-Qtr-Airl controls | 67/1471 | 63/1471 | 86/1471 | 62/1471 | 56/1471 | 36/1471 | 29/1471 | 21/1471 | 23/1471 |
| Adj R2 Statistic | 0.1487 [9] | 0.1895 [8] | 0.2430 [7] | 0.3010 [6] | 0.3456 [5] | 0.3724 [1] | 0.3649 [2] | 0.3549 [4] | 0.3566 [3] |
| AIC Statistic | 14,919 [2] | 24,402 [7] | 28,936 [9] | 26,041 [8] | 22,454 [6] | 18,641 [5] | 16,900 [4] | 15,002 [3] | 11,803 [1] |
| BIC Statistic | 19,947 [3] | 28,652 [7] | 34,121 [9] | 30,474 [8] | 26,176 [6] | 22,357 [5] | 20,236 [4] | 17,104 [2] | 14,225 [1] |
| RMSE Statistic | 0.2735 [1] | 0.3011 [2] | 0.3172 [8] | 0.3172 [8] | 0.3139 [7] | 0.3077 [4] | 0.3093 [5] | 0.3103 [6] | 0.3012 [3] |
| RMSE CV Statistic | 0.2957 [1] | 0.3357 [2] | 0.3638 [3] | 0.3759 [4] | 0.3859 [5] | 0.3912 [6] | 0.4052 [9] | 0.4044 [8] | 0.3983 [7] |
| Nr Obervations | 58,599 | 54,800 | 52,177 | 47,102 | 42,376 | 37,854 | 33,638 | 29,915 | 26,456 |

*Notes: Columns (1)-(9) estimated a fixed-effects, post-double-selection (PDS) LASSO-based methodology of Belloni et al. (2012, 2014a,b). LASSO penalty loadings account for city-pairs clustering. Post-LASSO estimation is performed with a fixed-effects procedure with standard errors robust to heteroskedasticity and autocorrelation. Control variables estimates omitted; "–" indicates that the LASSO procedure inactivated the variable; a number between square brackets denotes the rank of the statistic value according to each criterion–from the "best" to the "worse" across columns; ΔFLMOD, ASK, GROWTH SR, GROWTH LR, and AGE not penalized by LASSO; p-value representations: \*\*\*p<0.01, \*\* p<0.05, \* p<0.10.*